\documentclass[twocolumn,showpacs,preprintnumbers,amsmath,amssymb]{revtex4}

\usepackage{graphicx}
\usepackage{dcolumn}
\usepackage{bm}

\begin{document}

\title{Ultra narrow AuPd and Al wires}

\author{Fabio Altomare}
\author{Albert M. Chang}
\email{yingshe@phy.duke.edu}
\affiliation{%
  Physics Department, Duke University, Durham , NC 27707 \\
  Physics Department, Purdue University, West Lafayette, IN 47906 }

\author{Michael R. Melloch} \affiliation{ School of Electrical and
  Computer Engineering, Purdue University, West Lafayette, IN 47906 }

\author{Yuguang Hong}
\author{Charles W. Tu} 
\affiliation{ Department of Computer
  and Electrical Engineering, UCSD, La Jolla, CA 92093 }

\date{\today}

\begin{abstract}
In this letter we discuss a novel and versatile template technique aimed to the fabrication
of sub-10~nm wide wires. Using this technique, we have successfully measured AuPd wires, 12~nm wide 
and as long as 20 $\mu$m.  Even materials that form a strong 
superficial oxide, and thus not suited to be used in combination with 
other techniques, can be successfully employed.  In particular we have 
measured Al wires, with lateral width smaller or comparable to 
10~nm, and length exceeding 10~$\mu$m.
\end{abstract}


\maketitle

In recent years much effort has been devoted to the fabrication of
sub-10~nm wires, going beyond the capability of conventional nanofabrication 
techniques such as Electron Beam Lithograph (EBL), or Atomic Force Microscope 
lithography.  In the last few years, several groups have developed 
methods to reach such a limit.  Bezryadin et al. \cite{Bezryadin} have 
used carbon nanotubes as templates and the sputtering of metals onto 
nanotubes to form nanowires in order to study one-dimensional 
superconductivity.  Natelson et al.\cite{Nww+00} 
developed an edge technique, which employs selective etching of a 
GaAs layer, sandwiched between two AlGaAs layers, in order to define a 
trough, which is later filled by the evaporated metal.  Using this 
technique, they were able to fabricate Au$_{0.6}$Pd$_{0.4}$ (AuPd from 
now on) wires 5~nm in width and perform electrical measurements on them.  
Melosh et al.\cite{Santa_Barbara} have developed a pattern transfer
technique in which nanowires arrays, from deposition onto a 
molecular-beam-epitaxy (MBE) grown template, were transferred to a substrate 
and subsequently contacted electrically, allowing the authors to measure 
15~nm wide single wires, 20 $\mu$m long, and arrays of 8~nm wide wires.
In this letter we will describe the successful implementation of a 
technique, proposed several years ago\cite{altomare_aps} that,
employing an MBE template and EBL to define four-terminal measurement 
pads onto the template, allowed us to fabricate individual wires 7~nm wide 
and as long as 100~$\mu$m, and to make electrical measurements directly 
without the need of subsequent pattern transfer.  The nanowire is formed 
after a final metal evaporation onto a mechanical support that acts as a 
stencil; this support is provided by the (1$\bar{1}$0) plane of a MBE grown 
InP layer.  Because our method employs a single final evaporation to deposit 
the nanowire and to connect to the four-terminal measurement pads 
simultaneously, metals which form an oxide layer when exposed to air or 
oxygen, and therefore can be problematic to contact, such as aluminum, can 
readily be contacted.  Note that previously, successful fabrication 
and electrical measurement on sub-10~nm aluminum nanowires have not been reported.  
Our method combines the advantages offered by other techniques\cite{Bezryadin, Santa_Barbara, Nww+00}. Besides
the ease in electrical contact to different metals, e.g. AuPd, Al, 
additional advantages include: 
1)~the wire can be made extremely long: so far we
have been able to measure wire up to 100~$\mu$m\footnote{Results
  on such a wire will be presented elsewhere.}, 2)~the wire deposition 
is the last step in the process thus avoiding damage from subsequent processing,
3)~the absence of any resist on the template during 
the final deposition, enables materials requiring extremely
high or low temperature to be readily deposited, 4)~the width of the
wire is determined by the MBE growth and therefore atomically accurate and 
uniform, 5)~more complex, multiply connected geometries can in principle be 
fabricated with the appropriate MBE growth structure, 6)~the narrow stencil 
geometry restricts the size of the metal grains, enabling materials which do 
not wet well to form quality nanowires.
\begin{figure*}
 \includegraphics[width=\textwidth]{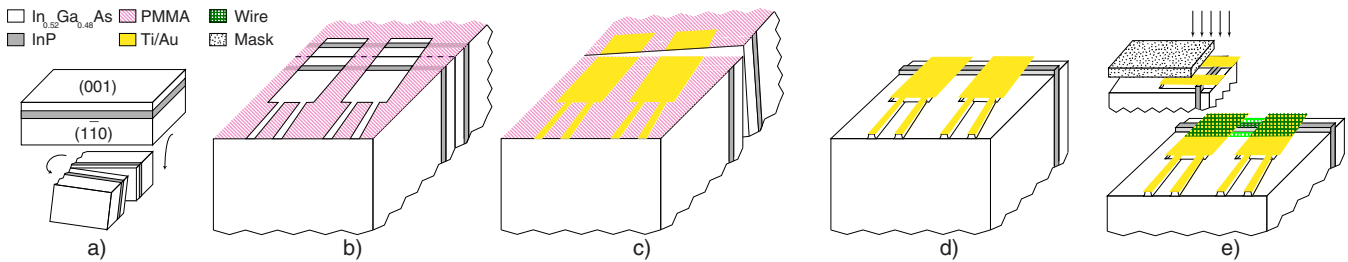}
    \caption{\label{fig:sample_fabrication}Schematic of sample fabrication: (a) The sample is cleaved
      in small strips which are cut in half and glued together with the two
      (001) plane facing each other, b) PMMA is spun on the
      (1$\bar{1}$0) crystallographic plane of the two pieces and a
      pattern is written using standard EBL and then developed, c)
      thermal evaporation is used to deposit a film of Ti/Au (the
      portion of the film deposited on the PMMA is not shown) and then
      the two halves are separated, d) after lift-off and oxygen plasma
      etching,  wet etching is used to define 
      the InP ridge, e) appropriately masking the substrate of the
      sample, the wire is formed through the final evaporation (top:
      side view of the evaporation arrangement, bottom:  final result).}
 \end{figure*}
\begin{table*}
\caption{\label{tab:table1}Sample parameters.}
\begin{minipage}[t]{\textwidth}
\begin{ruledtabular}
\begin{tabular}{ccccccccccc}
Sample &Material & w (nm) & t (nm) & L ($\mu$m) & R (k$\Omega$) & $\rho$
           ($\mu \Omega$cm)& $l_e$ (nm)    & D (c$m^2$/s)&  $L_{\phi}$
           (nm)   & $L_{so}$ (nm)\\\hline
au-1  &AuPd    & 42 & 12 &   5    & 5.59         & 55.8       &  1.5  & 7&
           27& --- \\
au-2 &AuPd    & 16 & 19&   10    & 23.3       & 70.6       &  1.2  & 5.6&
           38& --- \\
au-3  &AuPd\footnotemark[1]$^,$\footnotemark[2]    & 12 (8) & 8&   20    & 69.6       & 33.4 (22.3)      &  2.5 (3.7)  & 12 (17)&           81 (99)& --- \\
al-1 & Al\footnotemark[2]
& 6.9 (8) & 9.0 (12.5)& 10    & 8.3       & 5.1 (8.3)       &  7.6 (4.7)  & 51 (32) &  741 (649)& 115 ($<$5e-4)  \\
\end{tabular}
\end{ruledtabular}
\footnotetext[1]{The resistivity for this sample is smaller because of the much faster  evaporation rate with respect to other AuPd wires.}
\footnotetext[2]{The number in parenthesis are calculated using the nominal size of the wire; the others using the parameters obtained from the fit.}
\end{minipage}
\end{table*}
 
The MBE template is formed as follows.   
The starting point is an undoped (001) GaAs
 substrate on which In$_{x}$Ga$_{1-x}$As is graded from x=0 to x=0.52
 with a grading of 0.16/$\mu$m. At this final In concentration,
 In$_{0.52}$Ga$_{0.48}$As is lattice matched with InP: a layer of this
 semiconductor is then grown for a thickness $d$ and a cap layer of
 1.6~$\mu$m of In$_{0.52}$Ga$_{0.48}$As completes the growth.  This $d$ 
sets the width of the template and therefore the nanowire width.  From the
 wafer, small strips (approximately 12~mm wide) are cut and each
 strip is cleaved in half along its width along the (1$\bar{1}$0) 
cleavage plane. The two halves, after
 careful alignment, are glued  together with their top surfaces (the
 (001) plane) facing each other (Fig.~\ref{fig:sample_fabrication}a). 
 Polymethyl methachrilate (PMMA), 950K in molecular weight, is spun at
 5000 rpm onto the (1$\bar{1}$0) plane of the two pieces glued
 together and the resist is cured by
 baking it  for two hours in a conventional oven at  170~$^\circ$C.  
By gluing the pieces back to back, we ensure that the PMMA is uniform 
in the region of the InP layer, which resides only 1.6~$\mu$m from 
the (001) surface. Standard EBL is used to pattern the PMMA on one of 
the two pieces 
(Fig.~\ref{fig:sample_fabrication}b).  The pattern consists of openings 
for four measurement pads, with two leads for each pad, which will be 
used for four terminal electrical measurements.  The pads are completed by 
thermal evaporation of a bilayer of Au/Ti (50/60 \AA ~in thickness), where 
the titanium (Ti) is deposited first to promote adhesion, followed by lift-off 
in acetone with ultrasonic agitation.  It is necessary to separate the two 
halves (Fig.~\ref{fig:sample_fabrication}c) before lift-off, to minimize 
the risk of damaging the InP layer close to the (001) top surface.  
The pads have two purposes: they define an etch mask and provide, 
through the leads, a means to  electrically  contact the wire.  
Since the pads will be measured in series with the wires, they are designed 
to have a small resistance (of the order of 20-30~$\Omega$).  
Any organic residue is removed by  exposure to oxygen plasma for 15~s. 
 The last step of the stencil fabrication consists in room temperature
 wet etching  with a solution of H$_3$PO$_4$:H$_2$O$_2$:H$_2$O  which
 etches In$_{0.52}$Ga$_{0.48}$As and GaAs at a rate of $\approx$ 8~\AA /s and
 leaves InP virtually  untouched\cite{Inp_No_Oxide}. Except for the
 InP, the wet etching attacked all semiconductors not protected by
 the metal pads, while producing an undercut all around the metal
 pads. What remains is a ridge of InP 
 connecting two consecutive pads: their  separation will determine the
 wire length (Fig.~\ref{fig:sample_fabrication}d).
The evaporation of the desired material in a direction perpendicular
 to the (1$\bar{1}$0) crystallographic plane will form the wire
 on the InP ridge.  To achieve this,
 the substrate of the sample is masked so that only the topmost region 
(in the (001) direction of growth) of the sample in the vicinity of the 
InP ridge is exposed to a
 highly directional evaporation. The evaporated material will be
 deposited on the pads as well, seamlessly contacting the wire on the
 ridge (Fig.~\ref{fig:sample_fabrication}e). The vertical walls of the InP 
ridge and the undercut around the pads, provided by the wet chemical etching, 
 will prevent the formation of shorting paths through the metal deposited 
elsewhere in the etched regions.

To test our  technique and  to  demonstrate its versatility, we have
fabricated wires of two materials: AuPd is well
known\cite{jjlin,Nww+00}, has a grain size of few nanometer and it has been used to characterize wires
of comparable size\cite{Nww+00b}; Al is a material technologically relevant as
interconnects in circuits and after short exposure to air forms a
strong superficial oxide. 
We will discuss extensively a set of three AuPd wires
fabricated on the same chip and a single Al wire fabricated on an 8~nm wide stencil; results on
thicker wires (as wide as 42~nm) are reported in Table~\ref{tab:table1}.
 \begin{figure}
   \includegraphics[width=0.45\textwidth]{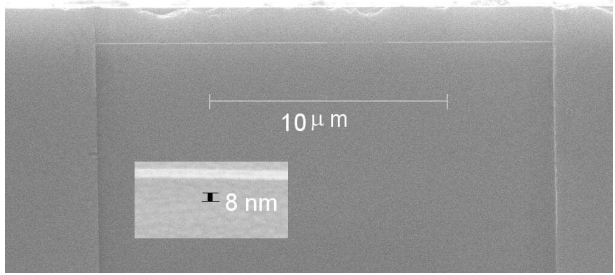}
  \caption{\label{8nm_SEM}SEM image of an 8~nm wide wire 20~$\mu$m long.} 
 \end{figure}
To distinguish the successful wire fabrication from accidental shorts
through the wide region evaporated (typically of the order of
20~$\mu$m)  we have performed magnetoresistance measurements in the weak
localization 
 regime which are sensitive to the dimensionality of the sample. 
 A sample is considered one-dimensional with respect to the weak 
 localization phenomenon if $L_{\phi} > w,t$ where $w$ and $t$
 represent the width and the thickness of the sample, and $L_{\phi}$ 
the quantum phase coherence length, respectively.  Theory for weak 
localization in 1D (1DWL) predicts\cite{echternach,pierre1} that the fractional 
change of the resistance ($\Delta R/R$) in an applied magnetic field depends 
on $R_{\square}$, $L_{\phi}$ and $L_{so}$ (the spin-orbit scattering 
length). Moreover, if  spin-orbit scattering is very strong, such as 
in AuPd, $\Delta R/R$ becomes independent on $L_{so}$ leaving $L_{\phi}$ 
as the only fitting parameter\cite{jjlin,Nww+00b}.

To detect the change in resistance, four terminal measurements were 
performed at 4.2~K, in magnetic fields up to 3 T, with a Princeton
Applied Research (PAR 124A) lockin-amplifier and an isolation
transformer to avoid possible ground loops in the measurement circuitry. 

\begin{figure}
  \includegraphics[width=0.45\textwidth]{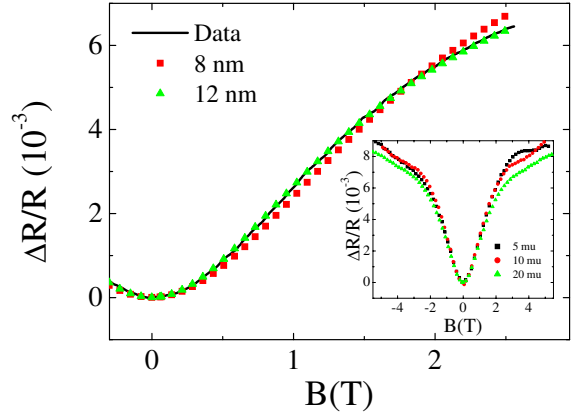}
\caption{\label{8nm_MR} 1DWL, at 4.2~K, for sample au-3 with fits to the 
    theory (see Tab.~\ref{tab:table1} for parameters). Inset: 
    This wire and two others fabricated on the same chip show a fractional 
    change in resistance almost identical as  expected from the theory.}
\end{figure}

Fig.~\ref{8nm_SEM} shows an SEM image of a typical 8~nm wide, 20~$\mu$m
long wire. The inset of  Fig.~\ref{8nm_MR} shows $\Delta R/R(H)$
for three 8~nm wide wires fabricated on the same chip 
by thermal evaporation of 8~nm of AuPd after the deposition of 1.5~nm 
of titanium as adhesion layer.  Their lengths are  5, 10 and 20~$\mu$m and 
their resistances are $R=20,\ 35.3\ ,69.6\ k\Omega$, respectively. 
Since, according to the weak localization theory, the fractional change 
of the resistance (in AuPd) depends only on $R_{\square}$ and $L_{\phi}$,
the inset of Fig.~\ref{8nm_MR} indicates our wire to be very uniform.  In
Fig.~\ref{8nm_MR} the data, together with a fit to the 1D theory of
weak localization, are shown for the 20~$\mu$m long wire. 

Au free electron model parameters\cite{asc-merm}, are combined with the 
measured resistivity to estimate the diffusion coefficient using
the Einstein relation \cite{jjlin,Nww+00b}.  For this set of 
wires, $\rho= 22.3~\mu \Omega cm$; the elastic mean free path $l_e=3.7~nm$ 
and the diffusion constant is $D \approx 1.7\times 10^{-3}~m^2/s$ 
The value of the quantum coherence length obtained by the fit 
($L_{\phi}=99.7~nm$) far exceeds the wire width.  This is a consistency check 
on the applicability of the 1D weak localization theory.
Moreover using the estimated diffusion constant, it is possible to
  calculate the dephasing scattering time 
$\tau _{\phi}=L_{\phi}^2/D=5.8\times 10^{-12}~s$ which is comparable to what
  obtained in other works. Treating the wire width as fitting parameter we 
obtain better agreement with $w=12~nm$ and $L_{\phi}=81~nm$ (Fig.~\ref{8nm_MR}).
\begin{figure}
  \includegraphics[width=0.45\textwidth]{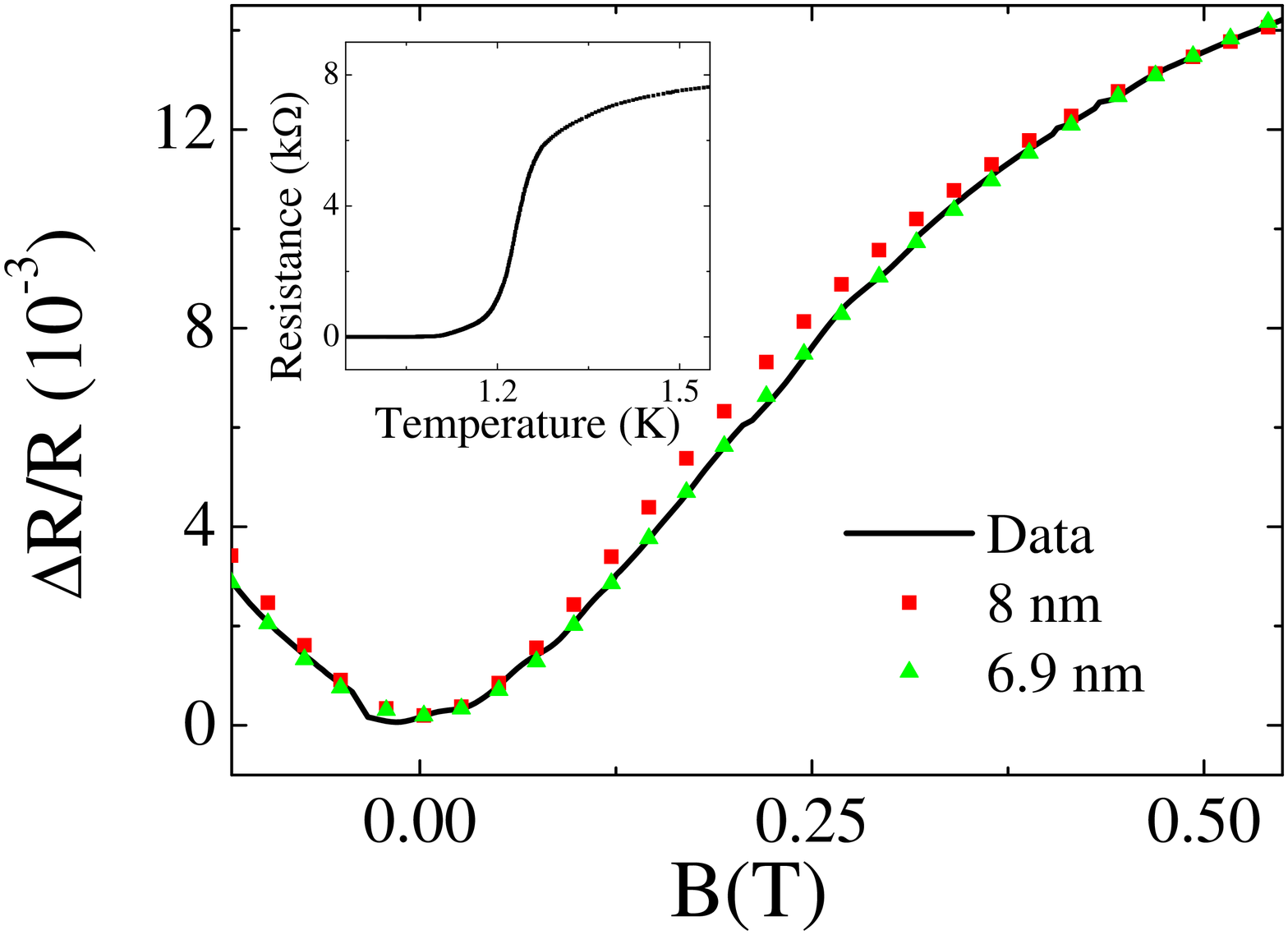}
  \caption{\label{8nm_Al_MR} 1DWL, at 4.2~K, for sample al-1 with fits to the 
    theory (see Tab.~\ref{tab:table1} for parameters). The inset shows 
    the superconducting transition of the wire.}
\end{figure}

 The aluminum wires have been fabricated by thermal evaporation from
 tungsten boat in a vacuum better than $7 \times 10^{-7}$ torr at an
 evaporation rate of about 10-12 \AA/s.  Titanium adhesion layer was 
 avoided to yield superior electrical characteristics.  
 We will discuss a single 8~nm wide wire, 12~nm thick and 10~$\mu$m long 
 ($R\approx8.3~k\Omega$ at T=4.2~K). 
 Fig.~\ref{8nm_Al_MR} shows the fractional change in resistance for this wire
with fits to the 1DWL. Since the mean free path is comparable to the wire width and thickness, the 1DWL dirty limit expression\cite{pierre1} is not appropriate and the clean limit expression has been used instead\cite{beenakker}.
Considering both the phase and the spin-orbit scattering length as  
fitting parameters, we obtain respectively $L_{\phi}= 666~nm$ and  $L_{so}<5\times 10^{-4}~nm$ using the nominal thickness of the wire. If we
 assume that the actual width of the wire is  smaller than the nominal one, due to oxidation effects, and we treat it
 as a fitting parameter we obtain a better agreement (Fig.~\ref{8nm_Al_MR})
 with  $L_{\phi}=741~nm$, $L_{so}=115~nm$ and $w=6.9~nm$. This would imply 
an  oxide thickness of about 3.5~nm, assuming the wire cross section to be 
similar to the AuPd one (after taking into account the different evaporation thickness).
Upon cooling at lower temperature the wire undergoes a superconducting
transition as shown in the inset of Fig.~\ref{8nm_Al_MR}.
The above examples illustrate the versatility and capabilities of our MBE-
template technique.  
The minimum wire width achievable with this technique is now under
investigation and the high yield in the fabrication of 
these wires (over 75\%) indicates that it may be possible to extend this
technique to the fabrication of even thinner wires.
As a final remark, note that the technique we have developed for 
InP and In$_{0.52}$Ga$_{0.48}$As can in principle be applied to other pairs 
of semiconductors as long as 1) a highly selective etching process
exist and 2) the semiconductor for the wire stencil 
does not oxidize. This work has been, in part, supported by NSF DMR-0135931 and DMR-0401648.


\begin{thebibliography}{11}
\expandafter\ifx\csname natexlab\endcsname\relax\def\natexlab#1{#1}\fi
\expandafter\ifx\csname bibnamefont\endcsname\relax
  \def\bibnamefont#1{#1}\fi
\expandafter\ifx\csname bibfnamefont\endcsname\relax
  \def\bibfnamefont#1{#1}\fi
\expandafter\ifx\csname citenamefont\endcsname\relax
  \def\citenamefont#1{#1}\fi
\expandafter\ifx\csname url\endcsname\relax
  \def\url#1{\texttt{#1}}\fi
\expandafter\ifx\csname urlprefix\endcsname\relax\def\urlprefix{URL }\fi
\providecommand{\bibinfo}[2]{#2}
\providecommand{\eprint}[2][]{\url{#2}}

\bibitem[{\citenamefont{Bezryadin et~al.}(2000)\citenamefont{Bezryadin, Lau,
  and Tinkham}}]{Bezryadin}
\bibinfo{author}{\bibfnamefont{A.}~\bibnamefont{Bezryadin}},
  \bibinfo{author}{\bibfnamefont{C.~N.} \bibnamefont{Lau}}, \bibnamefont{and}
  \bibinfo{author}{\bibfnamefont{M.}~\bibnamefont{Tinkham}},
  \bibinfo{journal}{Nature} \textbf{\bibinfo{volume}{404}},
  \bibinfo{pages}{971} (\bibinfo{year}{2000}).

\bibitem[{\citenamefont{Natelson
  et~al.}(2000{\natexlab{a}})\citenamefont{Natelson, Willett, West, and
  Pfeiffer}}]{Nww+00}
\bibinfo{author}{\bibfnamefont{D.}~\bibnamefont{Natelson}},
  \bibinfo{author}{\bibfnamefont{R.}~\bibnamefont{Willett}},
  \bibinfo{author}{\bibfnamefont{K.}~\bibnamefont{West}}, \bibnamefont{and}
  \bibinfo{author}{\bibfnamefont{L.}~\bibnamefont{Pfeiffer}},
  \bibinfo{journal}{Appl. Phys. Lett.} \textbf{\bibinfo{volume}{77}},
  \bibinfo{pages}{1991 } (\bibinfo{year}{2000}{\natexlab{a}}).

\bibitem[{\citenamefont{Melosh et~al.}(2003)\citenamefont{Melosh, Boukai,
  Diana, Gerardot, Badolato, Petroff, and Heath}}]{Santa_Barbara}
\bibinfo{author}{\bibfnamefont{N.}~\bibnamefont{Melosh}},
  \bibinfo{author}{\bibfnamefont{A.}~\bibnamefont{Boukai}},
  \bibinfo{author}{\bibfnamefont{F.}~\bibnamefont{Diana}},
  \bibinfo{author}{\bibfnamefont{B.}~\bibnamefont{Gerardot}},
  \bibinfo{author}{\bibfnamefont{A.}~\bibnamefont{Badolato}},
  \bibinfo{author}{\bibfnamefont{P.}~\bibnamefont{Petroff}}, \bibnamefont{and}
  \bibinfo{author}{\bibfnamefont{J.}~\bibnamefont{Heath}},
  \bibinfo{journal}{Science} \textbf{\bibinfo{volume}{300}},
  \bibinfo{pages}{112} (\bibinfo{year}{2003}).

\bibitem[{\citenamefont{Altomare et~al.}(2001)\citenamefont{Altomare, Chang,
  and Melloch}}]{altomare_aps}
\bibinfo{author}{\bibfnamefont{F.}~\bibnamefont{Altomare}},
  \bibinfo{author}{\bibfnamefont{A.~M.} \bibnamefont{Chang}}, \bibnamefont{and}
  \bibinfo{author}{\bibfnamefont{M.~R.} \bibnamefont{Melloch}}, in
  \emph{\bibinfo{booktitle}{Bull. Amer. Phys. Soc.}}
  (\bibinfo{organization}{APS}, \bibinfo{year}{2001}),
  vol.~\bibinfo{volume}{46}, pp. \bibinfo{pages}{866, {P}art II},
  \bibinfo{note}{abstract of the 2001 March Meeting S11-3.}

\bibitem[{\citenamefont{Pearsall}(2000)}]{Inp_No_Oxide}
\bibinfo{editor}{\bibfnamefont{T.}~\bibnamefont{Pearsall}}, ed.,
  \emph{\bibinfo{title}{Properties, Processing And Applications Of Indium
  Phospide}} (\bibinfo{publisher}{Inspec}, \bibinfo{year}{2000}),
  \bibinfo{note}{and reference therein}.

\bibitem[{\citenamefont{Lin and Giordano}(1987)}]{jjlin}
\bibinfo{author}{\bibfnamefont{J.}~\bibnamefont{Lin}} \bibnamefont{and}
  \bibinfo{author}{\bibfnamefont{N.}~\bibnamefont{Giordano}},
  \bibinfo{journal}{Phys. Rev. B} \textbf{\bibinfo{volume}{35}},
  \bibinfo{pages}{545} (\bibinfo{year}{1987}).

\bibitem[{\citenamefont{Natelson
  et~al.}(2000{\natexlab{b}})\citenamefont{Natelson, Willett, West, and
  Peiffer}}]{Nww+00b}
\bibinfo{author}{\bibfnamefont{D.}~\bibnamefont{Natelson}},
  \bibinfo{author}{\bibfnamefont{R.}~\bibnamefont{Willett}},
  \bibinfo{author}{\bibfnamefont{K.}~\bibnamefont{West}}, \bibnamefont{and}
  \bibinfo{author}{\bibfnamefont{L.}~\bibnamefont{Peiffer}},
  \bibinfo{journal}{Solid-State Commun.} \textbf{\bibinfo{volume}{115}},
  \bibinfo{pages}{269 } (\bibinfo{year}{2000}{\natexlab{b}}).

\bibitem[{\citenamefont{Echternach et~al.}(1993)\citenamefont{Echternach,
  Gershenson, Bozler, Bogdanov, and Nilsson}}]{echternach}
\bibinfo{author}{\bibfnamefont{P.~M.} \bibnamefont{Echternach}},
  \bibinfo{author}{\bibfnamefont{M.~E.} \bibnamefont{Gershenson}},
  \bibinfo{author}{\bibfnamefont{H.~M.} \bibnamefont{Bozler}},
  \bibinfo{author}{\bibfnamefont{A.~L.} \bibnamefont{Bogdanov}},
  \bibnamefont{and} \bibinfo{author}{\bibfnamefont{B.}~\bibnamefont{Nilsson}},
  \bibinfo{journal}{Phys. Rev. B} \textbf{\bibinfo{volume}{48}},
  \bibinfo{pages}{11516} (\bibinfo{year}{1993}).

\bibitem[{\citenamefont{Pierre et~al.}(2003)\citenamefont{Pierre, Gougam,
  Anthore, Pothier, Esteve, and Birge}}]{pierre1}
\bibinfo{author}{\bibfnamefont{F.}~\bibnamefont{Pierre}},
  \bibinfo{author}{\bibfnamefont{A.}~\bibnamefont{Gougam}},
  \bibinfo{author}{\bibfnamefont{A.}~\bibnamefont{Anthore}},
  \bibinfo{author}{\bibfnamefont{H.}~\bibnamefont{Pothier}},
  \bibinfo{author}{\bibfnamefont{D.}~\bibnamefont{Esteve}}, \bibnamefont{and}
  \bibinfo{author}{\bibfnamefont{N.}~\bibnamefont{Birge}},
  \bibinfo{journal}{Phys. Rev. B} \textbf{\bibinfo{volume}{68}},
  \bibinfo{pages}{085413} (\bibinfo{year}{2003}).

\bibitem[{\citenamefont{Ashcroft and Mermin}(1976)}]{asc-merm}
\bibinfo{author}{\bibfnamefont{N.~W.} \bibnamefont{Ashcroft}} \bibnamefont{and}
  \bibinfo{author}{\bibfnamefont{N.~D.} \bibnamefont{Mermin}},
  \emph{\bibinfo{title}{Solid State Physics}} (\bibinfo{publisher}{Saunders
  Colege Publishing}, \bibinfo{year}{1976}).

\bibitem[{\citenamefont{Beenakker and {van H}outen}(1988)}]{beenakker}
\bibinfo{author}{\bibfnamefont{C.~W.~J.} \bibnamefont{Beenakker}}
  \bibnamefont{and} \bibinfo{author}{\bibfnamefont{H.}~\bibnamefont{{van
  H}outen}}, \bibinfo{journal}{Phys. Rev. B} \textbf{\bibinfo{volume}{38}},
  \bibinfo{pages}{3232 } (\bibinfo{year}{1988}).

\end{thebibliography}
\end{document}